\documentclass[aps,prb,showpacs,twocolumn,superscriptaddress]{revtex4}
\usepackage{array,amsmath,amssymb,bm,graphicx}

\begin{document}

\title{Effect of spatial resolution on the estimates of the coherence length of
excitons in quantum wells}

\author{M. M. Fogler}
\author{Sen Yang}
\author{A. T. Hammack}
\author{L. V. Butov}
\affiliation{Department of Physics, University of California San Diego, La
Jolla, California 92093}

\author{A. C. Gossard}
\affiliation{Materials Department, University of California at Santa Barbara,
Santa Barbara, California 93106}
\date{\today}

\begin{abstract}

We evaluate the effect of diffraction-limited resolution of the optical system
on the estimates of the coherence length of two-dimensional excitons deduced
from the interferometric study of the exciton emission. The results are applied
for refining our earlier estimates of the coherence length of a cold gas of
indirect excitons in coupled quantum wells [S.~Yang \textit{et al.\/}, Phys.\
Rev.\ Lett.\ \textbf{97}, 187402(2006)]. We show that the apparent coherence
length is well approximated by the quadratic sum of the actual exciton coherence
length and the diffraction correction given by the conventional Abbe limit
divided by $\pi$. In practice, accounting for diffraction is necessary only when
the coherence length is smaller than about one wavelength. The earlier
conclusions regarding the strong enhancement of the exciton coherence length at
low temperatures remain intact.

\end{abstract}


\maketitle

\section{Introduction}
\label{sec:Introduction}

Spatial coherence of a bosonic system is encoded in its one-body density matrix
\begin{equation}\label{eq:rho}
\rho(\textbf{r}) = \langle \Psi^\dagger(\textbf{r}^\prime)
                 \Psi(\textbf{r}^\prime + \textbf{r}) \rangle\,,
\end{equation}
where $\Psi^\dagger$ ($\Psi$) is the particle creation (annihilation) operator
and the averaging is over both the quantum state and the position
$\textbf{r}^\prime$. In an isotropic system $\rho(\textbf{r})$ depends on the
absolute distance $r = |\textbf{r}|$ only. The density matrix decreases with $r$
as a result of scrambling the phases of the particles' wavefunctions by
scattering and thermal fluctuations. When this decrease is faster than $1 / r^2$,
we can define a characteristic decay length of $\rho(r)$ ---
the coherence length --- by the relation
\begin{equation}\label{eq:xi_x_def}
   \xi_x = \left({\displaystyle \int\limits_0^\infty \! \rho(r) r d r}\right)
   \!\Bigg{/}
   \left({\displaystyle \int\limits_0^\infty \! \rho(r) d r}\right)\,.
\end{equation}
Coherence length $\xi_x$ provides a quantitative information about
fundamental properties of the system of interest. For example, a rapid growth of
$\xi_x$ is anticipated as the bosons are cooled down below the temperature of
their quantum degeneracy.~\cite{Popov_83, Kagan_00} In addition,
$\xi_x$ sheds light on interactions and disorder in the system.

A gas of indirect excitons in GaAs coupled quantum wells is an example of a
solid-state system where this rich physics can be studied. A number of basic
physical parameters of such gases (concentration, exciton lifetime,
\textit{etc\/}.) can be controlled to bring them to a quasi-equilibrium at very
low temperatures.~\cite{Butov04r} The coherence of excitons is imprinted on the
coherence of the light they emit.~\cite{Ostereich, Laikhtman, Olaya-Castro,
Keeling_04, Zimmermann_05} This allows one to measure $\xi_x$ by optical
methods. In this paper we discuss a particular real-space technique, which has
proved to work in our experimental conditions.~\cite{Yang_06} It has enabled us
to determine the exciton coherence length $\xi_x$ and confirm its rapid
increase as temperature $T$ drops below a few degrees K.

Traditionally, real-space measurements of the optical coherence are done by
two-slit (or two-point) interferometry. However, this method becomes inaccurate
when $\xi_x$ is smaller than the spatial width of the regions from which the
light is collected. This is the case in our experiment where $\xi_x$ does not
exceed a few microns. However, our technique circumvents this limitation by using a
single pinhole. It works well in the regime $\xi_x < D / M_1$, where $D$ is the
pinhole diameter and $M_1$ is the image magnification factor. In other words,
the smallest measurable $\xi_x$ is determined not by $D / M_1$ but by the
finite spatial resolution of the optical system. In this paper we show how this
resolution can be taken into account.


The paper is organized as follows. In Sec.~\ref{sec:Results} we summarize the
main elements of the experimental technique and present our principal results.
In Sec.~\ref{sec:Summary} we review the theoretical model used in
Ref.~\onlinecite{Yang_06}. In Sec.~\ref{sec:Diffraction} we refine it to
incorporate the finite-resolution effects. Discussion and conclusions are given
in Sec.~\ref{sec:Conclusions}.

\section{Results}
\label{sec:Results}

\subsection{Measured exciton coherence length}
 
\begin{figure}[t]
\includegraphics[width=8.5cm]{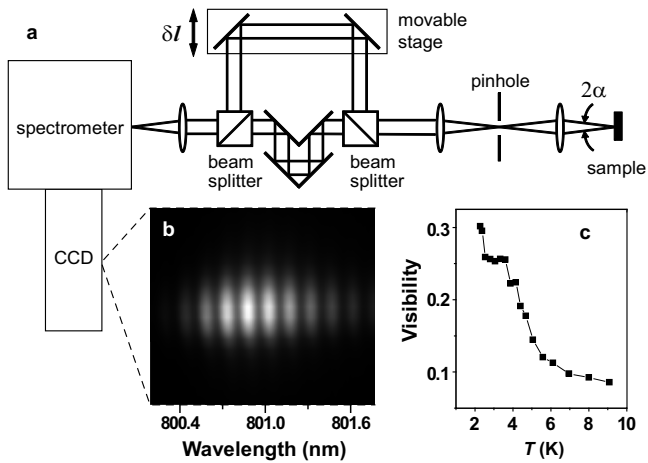}
\caption{ \label{fig:setup}
(a) Experimental setup. The collection angle of
the lens $2\alpha = 32^\circ$.
(b) The interference pattern on the CCD for $D = 25\,\mu\text{m}$,
$\delta l = 4.2\,\text{mm}$, and $T = 1.6\,\text{K}$.
(c) Visibility function $V(T)$ measured in
Ref.~\onlinecite{Yang_06} for $D = 50\,\mu\text{m}$ and $\delta l =
4.2\,\text{mm}$.
}
\end{figure}

In order to introduce the key parameters of the problem, we need to briefly
review the basic steps of the experimental implementation of our method for
measuring the coherence length.~\cite{Yang_06} The exciton PL is collected from
an area of size $D / M_1 = 2$--$10\,\mu\text{m}$ in the middle of one of the
exciton beads \cite{Butov02} ranging $\sim 30\,\mu\text{m}$ across. This is done
by placing a pinhole of diameter $D = 10$--$50\,\mu\text{m}$ at the intermediate
image plane of magnification $M_1 = 5$, see Fig. \ref{fig:setup}a. The light is
then passed through a Mach-Zehnder (MZ) interferometer with a tunable delay
length $\delta l$. The output of the interferometer is further magnified by the
factor $M_2 \approx 1.6$ (so that the total magnification factor is $M = M_1
M_2$) and then dispersed with a grating spectrometer, resulting in a
periodically modulated interference pattern. The intensity $I = I(x)$ of this
pattern is recorded by a CCD (Fig.~\ref{fig:setup}b). Here $x$ is the coordinate
along the CCD image. The visibility contrast, $V = (I_{\max} - I_{\min}) /
(I_{\max} + I_{\min})$, is calculated (Fig.~\ref{fig:setup}c). Finally, a
theoretical model that relates $V$ to $\xi_x$ is used to analyze the data and
determine $\xi_x$ as a function of temperature.

%
%
%

\begin{figure}[b]
\includegraphics[width=2.8in]{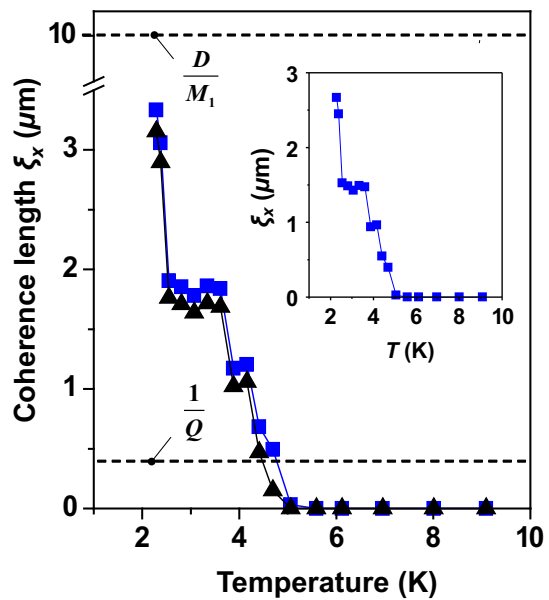}
\caption{ \label{fig:xi_x}
Main panel: $\xi_x$ deduced from
the fits of $V(T)$ to the theoretical curves in Fig.~\ref{fig:V}.
The triangles (squares) are the estimates with (without) taking into account the
spatial resolution of the experimental setup. The dashed lines indicate
boundaries of experimental accuracy. Inset: previous estimates of the coherence
length.~\cite{Serendipity} 
}
\end{figure}

The results of this analysis are shown in Fig.~\ref{fig:xi_x}. As
one can see, at $T < 4\,\text{K}$ the exciton coherence length grows to
a few $\mu\text{m}$, which exceeds the thermal de Broglie wavelength
\begin{equation}\label{eq:lambda_dB}
\lambda_\text{dB} = \left(\frac{2\pi \hbar^2}{m k_B T}\right)^{1 /
2}
\end{equation}
by an order of magnitude ($\lambda_\text{dB} \sim 0.1\, \mu\text{m}$ at $T =
2\,\text{K}$). Here $m = 0.2$ is the exciton mass in the quantum well in units
of the bare electron mass. The inequality $\xi_x \gg \lambda_\text{dB}$ is
anticipated for an exciton system near the superfluid transition. Possibility of
such a transition in systems with spatially separated electrons and holes has
been put forward in Refs.~\onlinecite{Lozovik_75, Shevchenko_76} and
\onlinecite{Fukuzawa_90}. The conventional estimate of the transition
temperature~\cite{Popov_83, Fisher_88, Prokofyev_02} $T_\text{BKT} \sim (\hbar^2
/ m) (n / g)$ gives a few degrees K for the exciton
concentrations~\cite{Comment_on_n} $n/g \sim 10^{10}\,\text{cm}^{-2}$ ($g = 4$
is the spin degeneracy).

%
%
%

In Ref.~\onlinecite{Yang_06} we used an approximation of geometrical
optics for describing the light collection in the apparatus. This is
justified in the most interesting regime of low $T$ where $\xi_x$ is
large. On the other hand, at the upper end of the temperature range
shown in Fig.~\ref{fig:xi_x} the estimated coherence length $\xi_x$ was
comparable to the diffraction-limited resolution of the optical
system, e.g., the Abbe limit~\cite{Abbe_1873,
Koehler_81}
\begin{equation}\label{eq:Abbe}
         \text{Ab} = \frac{\lambda_0}{2\, \text{NA}}\,.
\end{equation}
Here $\text{NA} = \sin \alpha$ is the numerical aperture and $\lambda_0$
is the wavelength. In our experiment, $\text{NA} = \sin 16^\circ \approx
0.3$ and $\lambda_0 \approx 800\,\text{nm}$.

In this paper we take diffraction into account and obtain refined estimates of
$\xi_x$, which are shown in Fig.~\ref{fig:xi_x} by the triangles. The difference
between previous and current estimates is insignificant for all but a few data
points at the boundary of the experimental resolution. Therefore, the case for a
rapid and strong onset of the spontaneous coherence of the exciton gas below the
temperature of a few degrees K remains intact.

\subsection{Relation between exciton and optical coherence lengths}

It is well understood that the Abbe limit is not a ``hard'' limit but
simply a characteristic measure of the optical resolution. In fact,
another commonly used formulas due to Rayleigh~\cite{Rayleigh,
Born_book} differ from Eq.~(\ref{eq:Abbe}) by numerical factors. Roughly
speaking, our theoretical model enables us to determine which numerical
factor is appropriate for our method of measuring $\xi_x$. More
precisely, our main result is as follows. Under certain assumptions,
$\xi_x$ is related to the \textit{optical\/} coherence length $\xi$ by
\begin{equation}\label{eq:xi_vs_xi_x}
\xi = \sqrt{\xi_x^2 + \frac{1}{Q^2}}\,\,\,,
\quad \frac{1}{Q} = \frac{\lambda_0}{2 \pi \text{NA}}\,.
\end{equation}
Here, in analogy to Eq.~(\ref{eq:xi_x_def}), $\xi$ is defined by
\begin{equation}\label{eq:xi_def}
   \xi = \frac{1}{M} \left({\displaystyle \int\limits_0^\infty \! g(0, R) R d R}\right)
   \!\Bigg{/}
   \left({\displaystyle \int\limits_0^\infty \! g(0, R) d R}\right)\,,
\end{equation}
where
\begin{equation}
g(t, R) = \langle E(t^\prime + t, \textbf{R}^\prime +
\textbf{R})
         E(t^\prime, \textbf{R}^\prime) \rangle /
        \langle E^2(t^\prime, \textbf{R}^\prime) \rangle
\label{eq:g}
\end{equation}
is the coherence function~\cite{Born_book} of the PL signal $E(t, \textbf{R})$
emitted by excitons and collected by the described system. In writing this
formula we assumed, for convenience, that the second magnification ($M_2$) of
the image occurs before the MZ interferometer, in which case $\textbf{R}$ and $R
= |\textbf{R}|$ are, respectively, the two-dimensional and the radial
coordinates in the plane of the fully magnified image.

Equation~(\ref{eq:xi_vs_xi_x}) is natural because an experimental
measurement of any length is affected by the spatial resolution limit.
However, Eq.~(\ref{eq:xi_vs_xi_x}) specifically indicates that for
$\xi_x$ measured using the setup depicted in Fig.~\ref{fig:xi_x}a, the
Abbe limit must be divided by $\pi$. This makes its effect
quantitatively smaller than one would naively think. 

\begin{figure}[t]
\includegraphics[width=2.8in]{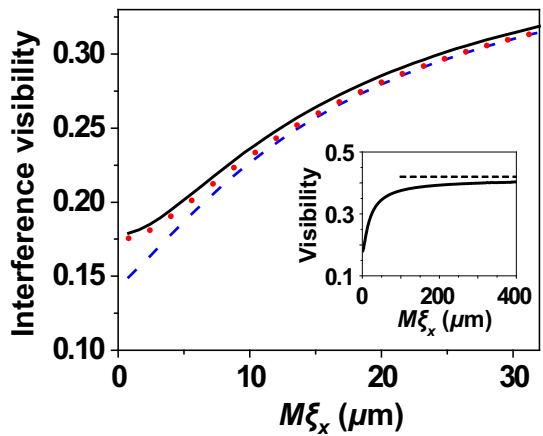}
\caption{\label{fig:V}
Visibility of the interference fringes \textit{vs.\/} $\xi_x$ for parameters
$\delta l = 4.2\,\text{mm}$, $D = 50\,\mu\text{m}$, $M_1 = 5$, $M_2 = 1.6$,
$Q^{-1} = 0.42\,\mu\text{m}$. The solid line is the current theory; the dashed
line is from Ref.~\onlinecite{Yang_06}; the dotted line is obtained from the
dashed one by the replacement $\xi_x \to \sqrt{\xi_x^2 + Q^{-2}}$. The inset shows
function $V(\xi_x)$ over a larger range of $\xi_x$, with the dashed
line indicating the asymptotic value $V(\xi_x = \infty)$.}
\end{figure}

In addition to the limitation from below, $\xi_x \gtrsim 1 / Q$, the
accuracy of the present method is restricted from above. When the
coherence length exceeds the size of the studied region of the sample $D
/ M_1$, the dependence of $V$ on $\xi_x$ should saturate, see the inset
of Fig.~\ref{fig:V}. This may become important at low enough
temperatures. The two limitations are indicated by the dashed lines in
Fig.~\ref{fig:xi_x}.

%
%

\section{Geometrical optics approach}
\label{sec:Summary}

In order to construct the model of the described above measurement
scheme, we need to know the functional form of $\rho(r)$. Unfortunately,
at present there is no comprehensive theoretical framework that provides
that. This is because $\rho(r)$ is affected by many factors, including
thermal broadening and a variety of scattering mechanisms, see more in
Sec.~\ref{sec:Conclusions}. On the other hand, the present state of
experiment~\cite{Yang_06} does not allow us to extract reliably anything
more than a characteristic decay length of function $\rho(r)$.
Therefore, we take a phenomenological approach and consider two simple
approximations for $\rho(r)$. The first one, used in
Ref.~\onlinecite{Yang_06}, is an exponential
\begin{equation}\label{eq:xi_x}
\rho(r) = \rho(0) \exp \left( -\frac{r}{\xi_x} \right)\,.
\end{equation}
This \textit{ansatz\/} should be reasonable over at least some range of
$r$ determined by the interplay of the disorder-limited mean-free path
($\sim 1\, \mu\text{m}$ in high-mobility GaAs structures), thermal
wavelength $\lambda_\text{dB} \sim 0.1\,\mu\text{m}$, and possibly, some
others. Most importantly, Eq.~(\ref{eq:xi_x}) provides a convenient
starting point because it contains a single characteristic length
$\xi_x$. Later in Sec.~\ref{sec:Diffraction} we will also consider a
Gaussian form,
\begin{equation}\label{eq:xi_x_Gauss}
\rho(r) = \rho(0) \exp \left( -\frac{r^2}{\pi \xi_x^2} \right)\,,
\end{equation}
for comparison.

The crucial point is the relation between $\rho(r)$ and $g(R)$. (Here
and below we drop $t$ in the argument of $g$ because the time-dependence
is not relevant for the calculation.) If the experimental apparatus can
be described by geometrical optics, then the only difference between the
two functions is the linear magnification $M$ and rescaling by a
constant factor. In this case, Eq.~(\ref{eq:xi_x}) entails
\begin{equation}\label{eq:xi}
g(R) = g(0) \exp \left( -{R} /\, {M \xi_x} \right)\,.
\end{equation}
In turn, Eq.~(\ref{eq:xi_def}) gives $\xi = \xi_x$, as expected. Until the very
end of this section we will use these two lengths interchangeably.

We have shown previously~\cite{Yang_06} that the interference visibility
contrast $V$ is related to $g$ as follows:
\begin{gather}
V = \theta(1 - \Delta) V_0\,,
\notag\\
V_0 = \frac{
          \displaystyle\int\limits_0^{1} \frac{d z}{z} \sin[F (1 - \Delta) z]
          \sin[F \Delta (1 - z)] g(z D_s)
         }
         {{F \Delta} \displaystyle \int\limits_0^{1} \frac{d z}{z} \sin(F z) (1 - z) g(z D_s)
         }\,,
\notag\\
F = \pi \frac{A N D_s}{\lambda_0},\quad
D_s = M_2 D,\quad
\Delta = \frac{\delta l}{N \lambda_0}\,,
\label{eq:V_general}
\end{gather}
where $\theta(z)$ is the step-function,~\cite{Comment_on_typos} $A$ is
the linear dispersion of the spectrometer, and $N$ is the number of
grooves in the diffraction grating. ($A = 1.55\,\text{nm}/\text{mm}$ and
$N = 1.5 \times 10^4$ in Ref.~\onlinecite{Yang_06}.)

For $g(r)$ given by Eq.~(\ref{eq:xi}) it is straightforward to compute the
integrals in Eq.~(\ref{eq:V_general}); however, in general it has to be done
numerically. For short coherence lengths, $\xi \ll \lambda_0 /\, A N M, D_s /
M$, one can also derive the analytical formula~\cite{Yang_06}
\begin{equation}\label{eq:V_small_xi}
    V \simeq \left(1 - \frac{\delta l}{N \lambda_0} \right)
                       \left|\frac{\sin \pi X}{\pi X} \right|\,,
\end{equation}
where
\begin{equation}\label{eq:delta_0}
         X = \frac{\delta l}{\delta l_0} \left(1 - \frac{M}{D_s}\, \xi \right)\,,\quad
\delta l_0 = \frac{\lambda_0^2}{A D_s}\,.
\end{equation}
This equation can be obtained by expanding the $\sin$-factors in the
integrals to the order $\mathcal{O}(z)$ and extending their integration limits
to infinity.

At $\delta l = \delta l_0$ and for small enough $\xi$, Eq.~(\ref{eq:V_small_xi}) yields
\begin{equation}\label{eq:V_near_node}
        V(\delta l_0, \xi) \simeq \left(1 - \frac{\lambda_0}{A N D_s} \right)
                                    \frac{M}{D_s}\, \xi\,.
\end{equation}
Thus, as such $\delta l$ the visibility contrast $V$ vanishes unless $\xi$ is
nonzero. Working with $\delta l \approx \delta l_0$ ensures the highest
sensitivity to $\xi$. In our experiment,~\cite{Yang_06} we extracted $\xi$ at
$\delta l = 4.2\,\text{mm}$, which is close but not exactly equal to $\delta l_0
= 5.2\,\text{mm}$. Therefore, we computed $V$ using the full
formula~(\ref{eq:V_general}). The results are shown by the dashed line in
Fig.~\ref{fig:V}. We fitted this theoretical curve $V(\xi)$ to the
experimentally measured $V(T)$ (Fig.~\ref{fig:setup}c) using $\xi =
\xi(T)$ as an adjustable parameter. In this manner we obtain the graph shown by
the squares in the main panel of Fig.~\ref{fig:xi_x}. We see that the exciton
coherence length exhibits a dramatic enhancement at $T < 4\,\text{K}$. On the
other hand, at $T \sim 4\,\text{K}$ this approach gives $\xi_x = \xi \sim
\lambda_0$. One can anticipate that the geometrical optics becomes inaccurate at
such small $\xi$, so that $\xi$ and $\xi_x$ are in fact different. This question
is studied in the next section.

\section{Diffraction effects}
\label{sec:Diffraction}

%
%
%
\begin{figure}[b]
\includegraphics[width=2.5in]{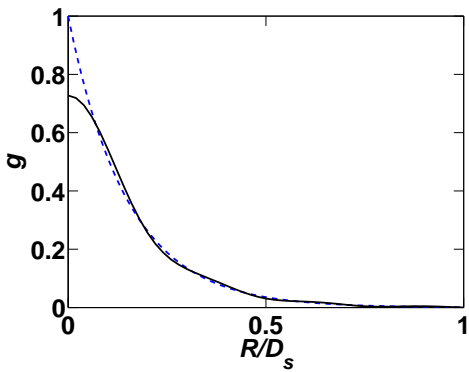}
\caption{ \label{fig:G} Optical coherence function $g = g(R)$. The solid line
is computed for the following set of parameters: $\xi_x = 1.5\,\mu\text{m}$, $M
= 8$, $Q^{-1} = 0.42\,\mu\text{m}$, and $D_s = 80\,\mu\text{m}$. The dashed line
is for geometrical optics, $Q^{-1} = 0$.}
\end{figure}
%

The conventional theory~\cite{Born_book} of the image formation in
optical instruments laid down by Abbe~\cite{Abbe_1873} in 1873
predicts that a point source imaged by a lens with a magnification $M$ creates
the diffraction spot
\begin{equation}\label{eq:Airy}
E(R) \propto \int\limits_{k < Q} \frac{d^2 k}{(2 \pi)^2}
       e^{i \textbf{kR} /{M}}
       = \frac{2 M Q}{R}\, J_1 \!\left(\frac{Q R}{M}\right)
\end{equation}
in the image plane. Here $R$ is the radial distance, $Q$ is given by
Eq.~(\ref{eq:xi_vs_xi_x}), and $J_1(z)$ is the Bessel function. The field
distribution~(\ref{eq:Airy}) is known as the Airy diffraction
pattern.~\cite{Airy_1841} The physical meaning of $Q$ is the largest tangential
wavenumber admitted by the lens. Accordingly, the diffraction can be
alternatively viewed as a low-pass filtering of the incoming light by the
lens.~\cite{Koehler_81}

The Airy pattern plays the role of the response function of the lens. Its finite
spread in $R$ imposes the limit on the achievable optical resolution $\sim
Q^{-1}$ and is the source of the difference between the optical and the actual
exciton coherence lengths, see Eq.~(\ref{eq:xi_vs_xi_x}). Indeed, because of the
diffraction, the coherence function $g(R)$ is not just a rescaled copy of
$\rho(R / M)$ but its convolution with the Airy pattern. Using tilde to denote
the 2D Fourier transform, we can express this fact as follows:
\begin{equation}\label{eq:g_from_rho}
\tilde{g}(k) \propto \theta(Q - M k) \tilde\rho(M k)\,.
\end{equation}
Note that $\tilde\rho(k)$ has the physical meaning of the momentum distribution
function for excitons. Computing $\tilde\rho(k)$ from Eq.~(\ref{eq:xi_x}), we
get
\begin{equation}\label{eq:g_new}
\tilde{g}(k) = \text{const}
          \times \frac{\theta(Q - M k)}{(1 + M^2 \xi_x^2 k^2)^{3 / 2}}\,.
\end{equation}
The constant prefactor in this formula has no effect on $V$. It is convenient to
choose it to be $2\pi M^2 \xi_x^2$, so that
\begin{equation}\label{eq:G_new}
g(R) = M^2 \xi_x^2 \int\limits_0^{Q/M}
       \frac{ J_0(k R) k d k}{(1 + M^2 \xi_x^2 k^2)^{3 / 2}}\,.
\end{equation}
In this case in the limit $Q \to \infty$, we recover Eq.~(\ref{eq:xi}) with
$g(0) = 1$. On the other hand, for finite $Q$, we have
\begin{equation}\label{eq:G_0}
g(0) = 1 - \frac{1}{\sqrt{1 + Q^2 \xi_x^2}}\,.
\end{equation}
Additionally, at large $R$, function $g(R)$ acquires the behavior
characteristic of the Airy pattern~(\ref{eq:Airy}): quasiperiodic oscillations
with the envelope decaying as $R^{-3/2}$, see Fig.~\ref{fig:G}. Finally,
computing the optical coherence length $\xi$ according to Eq.~(\ref{eq:xi_def})
we get Eq.~(\ref{eq:xi_vs_xi_x}).

The refined theoretical dependence of $V$ on $\xi_x$ can now be obtained by
substituting Eq.~(\ref{eq:G_new}) into Eq.~(\ref{eq:V_general}). As before, for
small $\xi_x$ analytical formulas (\ref{eq:V_small_xi})--(\ref{eq:V_near_node})
suffice, with $\xi$ defined by Eq.~(\ref{eq:xi_vs_xi_x}). When this $\xi$
becomes comparable to $D_s / M$, numerical evaluation of
Eqs.~(\ref{eq:V_general}) and (\ref{eq:G_new}) is necessary. The representative
results are shown by the solid line in Fig.~\ref{fig:V}. For comparison, two
other curves are included. The dashed line is the geometrical optics
approximation, $\xi = \xi_x$ of Sec.~\ref{sec:Summary}. The dotted line is the
result of correcting the latter according to Eq.~(\ref{eq:xi_vs_xi_x}) and using
$Q$ appropriate for our experiment. As one can see, at small $\xi_x$, the
effect of the diffraction-limited resolution of the optical system is indeed
accounted for by Eq.~(\ref{eq:xi_vs_xi_x}). At large $\xi_x$, the correction
becomes small and all the curves are very close to each other.

\begin{figure}[t]
\includegraphics[width=2.9in]{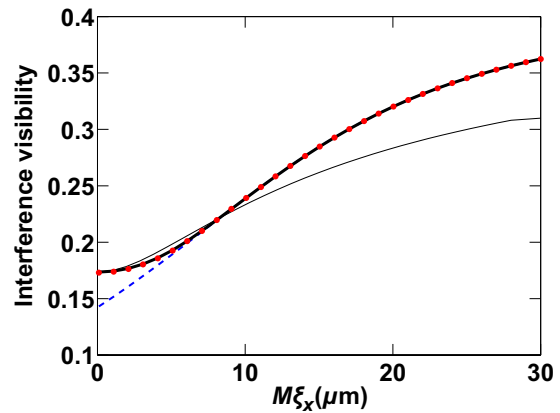}
\caption{\label{fig:V_Gauss}
Visibility of the interference fringes \textit{vs.\/} $\xi_x$ for the Gaussian
coherence function~(\ref{eq:xi_x_Gauss}) and the parameters of Fig.~\ref{fig:V}.
The thick solid line includes the diffraction correction; the dashed line is for the
geometrical optics; the dotted line is obtained from the dashed one by the
replacement of $\xi_x$ by $\xi$ according to Eq.~(\ref{eq:xi_vs_xi_x_Gauss}).
The thin solid line is $V(\xi_x)$ for the exponential \textit{ansatz\/},
replotted from Fig.~\ref{fig:V} to facilitate the comparison.
}
\end{figure}

It is instructive to examine how our conclusions so far depend on the model
assumption~(\ref{eq:xi_x}) about function $\rho(r)$. To this end we consider
next the Gaussian \textit{ansatz\/}~(\ref{eq:xi_x_Gauss}), which is similar to
Maxwell-Boltzmann distribution except the coefficient in the exponential factor
is adjusted to satisfy Eq.~(\ref{eq:xi_x_def}). Let us compute the corresponding
$\xi$. Using Eq.~(\ref{eq:g_from_rho}), we can rewrite Eq.~(\ref{eq:xi_def}) as
\begin{equation}\label{eq:xi_general}
  \xi  = {\tilde{\rho}(0)} \,\Big/\,
         {\left( \textstyle\int_0^Q d k \tilde{\rho}(k) \right)}\,.
\end{equation}
Substituting here
\begin{equation}\label{eq:xi_x_Gauss_q}
\tilde\rho(k) = \tilde\rho(0)
                \exp \left( -\frac{\pi}{4} k^2 \xi_x^2 \right)\,,
\end{equation}
which follows from Eq.~(\ref{eq:xi_x_Gauss}), we get
\begin{equation}\label{eq:xi_vs_xi_x_Gauss}
\xi = \frac{\xi_x}{\text{erf}\,[(\sqrt{\pi}/2) Q \xi_x]}\,,
\end{equation}
where $\text{erf}\,(z)$ is the error function. This formula replaces
Eq.~(\ref{eq:xi_vs_xi_x}). Interestingly, it implies that for the same $\xi_x$
and $Q$, the effect of the finite resolution in the case of a Gaussian decay is
always smaller than for the exponential one. The direct numerical evaluation of
Eq.~(\ref{eq:V_general}) with the Gaussian profile~(\ref{eq:xi_x_Gauss})
confirms this expectation, see Fig.~\ref{fig:V_Gauss}. Thus, we again conclude
that the diffraction correction is important for $\xi \lesssim \lambda_0$, but
it is very small in the most interesting region $\xi > \lambda_0$.

%
%

Notice that function $V(\xi_x)$ increases somewhat faster with $\xi_x$ for the
Gaussian case compared to the exponential one, cf.~Figs.~\ref{fig:V}
\textit{vs.\/} \ref{fig:V_Gauss}. Therefore, had we adopted the Gaussian
\textit{ansatz\/}~(\ref{eq:xi_x_Gauss}), the deduced values of $\xi_x(T)$ would
have been somewhat smaller than those plotted in Fig.~\ref{fig:xi_x}. This is to
be expected: if the exact functional form of $\rho(r)$ is unknown, its
characteristic decay length can be determined only up to a numerical coefficient
of the order of unity.

\section{Discussion}
\label{sec:Conclusions}

The main purpose of the present work is refinement of the optical method for
determining the exciton coherence length $\xi_x$. Unlike previously proposed
schemes,~\cite{Keeling_04, Zimmermann_05} which involve angle-resolved
photoluminescence, our technique is based on real-space interferometry.

We showed that in order to obtain an accurate estimate of $\xi_x$, the
optical coherence length $\xi$ of the exciton emission should be corrected
because of the diffraction-limited spatial resolution of the experimental
apparatus. However the correction is insignificant as long as $\xi_x$ is larger
than about one wavelength and the numerical aperture $\text{NA}$ of the
experimental setup is not too small. The correction does grow as $\text{NA}$
decreases, and so reduced $\text{NA}$ should be avoided.

It is well known~\cite{Abbe_1873, Koehler_81, Born_book} that limitation of the
spatial resolution due to diffraction is equivalent to that due to restriction
on tangential wavenumbers $k$ admitted by the lens collecting the signal. This
$k$-filtering effect has been considered in Ref.~\onlinecite{Ivanov_xxx} in
application to the measurements of the exciton coherence length. For the
collection angle $\alpha = 16^\circ$ in our experiments,~\cite{Yang_06} the
results presented in Fig.~3c of Ref.~\onlinecite{Ivanov_xxx} give the
correlation length due to the $k$-filtering effect $\xi_{\gamma} \approx
1\,\mu\text{m}$. (This length plays the role similar to that of $Q^{-1} =
0.42\,\mu\text{m}$ in our formalism.) For the considered $\rho(r)$ this
correction enters either through the quadratic sum, Eq.~(\ref{eq:xi_vs_xi_x}),
or the error function, Eq.~(\ref{eq:xi_vs_xi_x_Gauss}). As a result, the
estimation of the $k$-filtering effect per Ref.~\onlinecite{Ivanov_xxx} gives
only a small ($\sim 10\%$) correction, e.g., $\xi - \sqrt{\xi^2 -
\xi_{\gamma}^2}$ to the optical coherence length $\xi$ measured in
Ref.~\onlinecite{Yang_06} at low $T$. Therefore, it cannot explain the observed
large enhancement of the coherence length at $T < 4\,\text{K}$. Our calculations
indicate that the correction is even smaller.

The discussion of physics that is responsible for the observed rapid change in
$\xi_x$ at low temperatures is however beyond the score of this paper. As a
final word, we would like to offer only the following minimal remarks on this
matter.

The density matrix $\rho(r)$, from which $\xi_x$ is defined through
Eq.~(\ref{eq:xi_x_def}), is influenced by a number of factors, including Bose
statistics, interactions, and scattering. The effect of the first two has been
studied extensively, albeit for simplified models of interaction. According to
present understanding,~\cite{Popov_83, Kagan_00} the long-distance behavior of
function $\rho(r)$ is qualitatively different above and below the
Berezinskii-Kosterlitz-Thouless (BKT) transition temperature $T_\text{BKT}$. At
$T \gg T_\text{BKT}$, where $\tilde\rho(k)$ obeys the classical Boltzmann
statistics, $\rho(r)$ decays as a Gaussian, Eq.~(\ref{eq:xi_x_Gauss}), with the
coherence length
\begin{equation}\label{eq:g_Gauss}
             \xi_x = \lambda_\text{dB} / \pi\,.
\end{equation}
[Our estimates of $\xi_x$ at $T < 4\,\text{K}$
exceed $\lambda_\text{dB} / \pi$ by an order of magnitude, suggesting that
Eq.~(\ref{eq:g_Gauss}) does not apply at such temperatures.] At $T
< T_\text{BKT}$, the eventual asymptotic decay of the density matrix becomes
algebraic, $\rho(r) \propto r^{-\nu}$ with a temperature-dependent exponent
$\nu(T)$. The behavior of $\rho(r)$ at intermediate temperatures and/or
distances is more complicated. In general, it can be computed only numerically,
e.g., by quantum Monte-Carlo method.~\cite{Kagan_00}

Some of the other mechanisms of limiting the coherence length $\xi_x$, such as
exciton recombination and exciton-phonon scattering are too weak to
significantly affect the large magnitude of observed $\xi_x$ at low
temperatures.~\cite{Yang_06} However, scattering by impurities and defects
should be seriously considered. It can substantially modify the functional form
of $\rho(r)$ compared to the disorder-free case. Indeed, weak disorder typically
leads to an exponential decay of the correlation functions on the scale of the
mean-free path, which in fact inspired our \textit{ansatz\/}~(\ref{eq:xi_x}). As
temperature goes down, the strength of the disorder decreases because excitons
can screen it more efficiently.~\cite{Butov98, Nikonov98, Ivanov02} This should
increase both the mean-free path and the exciton coherence length.

The comprehensive theoretical calculation of the exciton coherence length that
would take into account all relevant thermal, interaction, and disorder
screening effects is yet unavailable.

\acknowledgements

We thank A.~Ivanov and L.~Mouchliadis for valuable discussions and
comments on the manuscript. This work is supported by NSF grants
DMR-0606543, DMR-0706654, and ARO grant W911NF-05-1-0527.



\begin{thebibliography}{99}
\bibitem{Serendipity} Here we remedy an error in handling parameter
$M_2$ in the previous paper.~\cite{Yang_06} Therein $M_2 = 1.6$ was
determined from the data analysis. However, by confusion the estimated
value of $M\xi$ was divided by $M = 10$ (instead of $M = 5 \times 1.6 =
8$) to give the result for $\xi$. Accordingly, the values of $\xi$ in
Fig.~3e of Ref.~\onlinecite{Yang_06} are somewhat smaller than what
should have been shown. (For ease of comparison, this figure is
reproduced in the inset of Fig.~\ref{fig:xi_x}.) In the main panel
of Fig.~\ref{fig:xi_x} and the rest of the present paper this error has
been corrected: we use $M_2 = 1.6$ and $M = 8$ throughout.
\bibitem{Comment_on_n} Exciton concentration $n$ can be deduced from the
exciton energy $E_x = h c / \lambda_0$ using the linear relation $n = (C / e^2)
[E_x - E_x(n = 0)]$. However, the value of the coefficient $C$ is presently
under debate. Earlier literature~\cite{Yoshioka90, Zhu95, Lozovik96,
Ivanov02} assumed that $C$ is equal to the classical capacitance per
unit area $C = \epsilon / (4 \pi d)$, where $\epsilon$ is the
dielectric constant and $d$ is the separation between the electron and
hole layers. A more recent calculation~\cite{Zimmermann08} gives an
order of magnitude higher estimate for $C$ and therefore $n$ in our
experiments.
\bibitem{Comment_on_typos} The absence of ``$\theta$'' and the extra
``$\sin$'' in the corresponding Eq.~(7) of Ref.~\onlinecite{Yang_06} are
typographic errors.
\end{thebibliography}
\end{document}